\documentclass[8.5pt,twoside,twocolumn]{article}
\usepackage[super,sort&compress,comma]{natbib} 
\usepackage{mhchem}
\usepackage{times,mathptmx}
\usepackage{sectsty}
\usepackage{balance} 

\usepackage{graphicx} 
\usepackage{lastpage}
\usepackage[format=plain,justification=raggedright,singlelinecheck=false,font=small,labelfont=bf,labelsep=space]{caption} 
\usepackage{fancyhdr}
\usepackage{url}
\newcommand{\R}{\mathcal{R}}
\renewcommand{\S}{\mathcal{S}}

\begin{document}

\thispagestyle{plain}
\fancypagestyle{plain}{
\fancyhead[L]{\includegraphics[height=8pt]{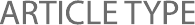}}
\fancyhead[C]{\hspace{-1cm}\includegraphics[height=20pt]{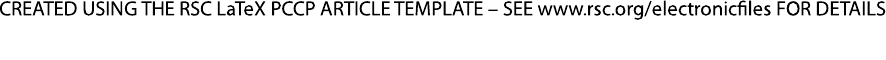}}
\fancyhead[R]{\includegraphics[height=10pt]{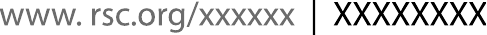}\vspace{-0.2cm}}
\renewcommand{\headrulewidth}{1pt}}
\renewcommand{\thefootnote}{\fnsymbol{footnote}}
\renewcommand\footnoterule{\vspace*{1pt}%
\hrule width 3.4in height 0.4pt \vspace*{5pt}} 
\setcounter{secnumdepth}{5}

\makeatletter 
\def\subsubsection{\@startsection{subsubsection}{3}{10pt}{-1.25ex plus -1ex minus -.1ex}{0ex plus 0ex}{\normalsize\bf}} 
\def\paragraph{\@startsection{paragraph}{4}{10pt}{-1.25ex plus -1ex minus -.1ex}{0ex plus 0ex}{\normalsize\textit}} 
\renewcommand\@biblabel[1]{#1}            
\renewcommand\@makefntext[1]%
{\noindent\makebox[0pt][r]{\@thefnmark\,}#1}
\makeatother 
\renewcommand{\figurename}{\small{Fig.}~}
\sectionfont{\large}
\subsectionfont{\normalsize} 

\fancyfoot{}
\fancyfoot[LO,RE]{\vspace{-7pt}\includegraphics[height=9pt]{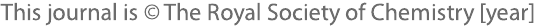}}
\fancyfoot[CO]{\vspace{-7.2pt}\hspace{12.2cm}\includegraphics{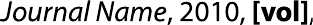}}
\fancyfoot[CE]{\vspace{-7.5pt}\hspace{-13.5cm}\includegraphics{headers/RF}}
\fancyfoot[RO]{\footnotesize{\sffamily{1--\pageref{LastPage} ~\textbar  \hspace{2pt}\thepage}}}
\fancyfoot[LE]{\footnotesize{\sffamily{\thepage~\textbar\hspace{3.45cm} 1--\pageref{LastPage}}}}
\fancyhead{}
\renewcommand{\headrulewidth}{1pt} 
\renewcommand{\footrulewidth}{1pt}
\setlength{\arrayrulewidth}{1pt}
\setlength{\columnsep}{6.5mm}
\setlength\bibsep{1pt}

\twocolumn[
  \begin{@twocolumnfalse}
\noindent\LARGE{\textbf{Enrichment and aggregation of topological motifs are
    independent organizational principles of integrated interaction networks$^\dag$}}
\vspace{0.6cm}

\noindent\large{\textbf{Tom Michoel,$^{\ast}$\textit{$^{a}$} Anagha
    Joshi,\textit{$^{b}$} Bruno Nachtergaele,\textit{$^{c}$} and Yves
    Van de Peer\textit{$^{d}$}}}\vspace{0.5cm}

\noindent\textit{\small{\textbf{Received Xth XXXXXXXXXX 20XX, Accepted Xth XXXXXXXXX 20XX\newline
First published on the web Xth XXXXXXXXXX 200X}}}

\noindent \textbf{\small{DOI: 10.1039/b000000x}}
\vspace{0.6cm}

\noindent \normalsize{Topological network motifs represent functional
  relationships within and between regulatory and protein-protein
  interaction networks.  Enriched motifs often aggregate into
  self-contained units forming functional modules.  Theoretical models
  for network evolution by duplication-divergence mechanisms and for
  network topology by hierarchical scale-free networks have suggested
  a one-to-one relation between network motif enrichment and
  aggregation, but this relation has never been tested quantitatively
  in real biological interaction networks. Here we introduce a novel
  method for assessing the statistical significance of network motif
  aggregation and for identifying clusters of overlapping network
  motifs. Using an integrated network of transcriptional,
  posttranslational and protein-protein interactions in yeast we show
  that network motif aggregation reflects a local modularity property
  which is independent of network motif enrichment. In particular our
  method identified novel functional network themes for a set of
  motifs which are not enriched yet aggregate significantly and
  challenges the conventional view that network motif enrichment is
  the most basic organizational principle of complex networks.}  
\vspace{0.5cm}
 \end{@twocolumnfalse}
  ]

\section{Introduction}
\footnotetext{\dag~Electronic Supplementary Information (ESI)
  available: [details of any supplementary information available
  should be included here]. See DOI: 10.1039/b000000x/}


\footnotetext{\textit{$^{a}$~Freiburg Institute for Advanced Studies
    (FRIAS), University of Freiburg, Albertstrasse 19, 79104 Freiburg,
    Germany. Fax: +49 761 203 97323; Tel: +49 761 203 97346; E-mail:
    tom.michoel@frias.uni-freiburg.de}}
\footnotetext{\textit{$^{b}$~Department of Haematology, Cambridge
    Institute for Medical Research, University of Cambridge, Wellcome
    Trust/MRC Building Hills Rd, Cambridge CB2 0XY, United Kingdom. }}
\footnotetext{\textit{$^{c}$~Department of Mathematics, University of
    California, Davis, One Shields Avenue, Davis, CA 95616-8366, USA}}
\footnotetext{\textit{$^{d}$~Department of Plant Systems Biology, VIB
    and Department of Plant Biotechnology and Genetics, Ghent
    University, Technologiepark 927, B-9052 Gent, Belgium}}



Reconstructing the organizational principles that determine the
structure and function of regulatory and protein-protein interaction
networks is a key challenge of network biology. Network motifs, small
subgraphs occuring significantly more often than expected by chance,
have been proposed as the basic building blocks of complex networks
\cite{milo2002,shen-orr2002}, including integrated networks composed
of multiple types of interactions
\cite{yeger-lotem2004,zhang2005,ptacek2005,yu2006,fiedler2009}. In
transcriptional regulatory networks, network motifs are known to
aggregate into larger, self-contained units
\cite{kashtan2004,shen-orr2002,dobrin2004}.  This concept was extended
to integrated networks and resulted in the definition of `network
themes', frequently recurring higher-level patterns of overlapping
network motifs \cite{zhang2005}, which characterize the structure of
functional modules \cite{hartwell1999}. Several studies have further
investigated the connection between network motif enrichment and
aggregation, from a topological as well as from an evolutionary
perspective.  In hierarchical scale-free random networks the
enrichment and aggregation of a certain class of subgraphs are
intimately related to each other and to the global topological network
parameters \cite{vazquez2004}.  Furthermore these subgraphs tend to
aggregate around network hubs \cite{vazquez2004}.  A comparative
phylogenetic analysis of genes within motifs has shown that they are
not subject to any evolutionary pressure to preserve the motif pattern
\cite{mazurie2005}. A likely reason is that the motifs aggregate and
cannot be considered in isolation \cite{mazurie2005,sole2006}. In a
simple duplication-divergence model for network growth, modularity,
accompanied by subgraph abundance, can appear for free without
selection pressure \cite{sole2008}. On the other hand, in a model of
evolving electronic circuits, modularity emerges only in an
environment that changes itself in a modular manner, while network
motif enrichment appears only if the modularly varying goal contains
information-processing tasks \cite{kashtan2005}.

An important question that has not been addressed before is whether
the aggregation of a motif is indeed surprising or significant, given
the number of motif instances that have to fit on a network with a
fixed degree distribution, and if so, how such aggregation relates to
motif enrichment and whether any functional interpretation can be
given to it. Here we address this question using random network
ensembles which preserve the degree distribution as well as the total
motif count, a new network motif aggregation statistic, and a novel
algorithm for identifying clusters of overlapping motifs to assess in
a quantitative way the enrichment and aggregation significance of all
composite motifs in a network which integrates transcriptional
\cite{harbison2004} and posttranslational \cite{ptacek2005} regulatory
interactions as well as physical protein-protein interactions
\cite{stark2006} in yeast.

\section{Results}
\label{sec:results}

\subsection{Molecular interaction networks in yeast deviate from the
  hierarchical scale-free model}
\label{sec:molec-inter-netw}

The most detailed study of the relation between network motif
enrichment and aggregation to date has been done for so-called
hierarchical scale-free random networks, which are characterized by a
power-law degree distribution $P(k)\sim k^{-\gamma}$, where $P(k)$ is
the probability for a node to have $k$ neighbors (irrespective of edge
direction), and a power-law scaling for the clustering coefficient
$C(k)\sim k^{-\alpha}$ where $C(k)$ is the average clustering
coefficient for a node with $k$ neighbors
\cite{vazquez2004}. Hierarchical scale-free random networks share with
biological networks the property that they are organized into many
small, highly connected modules that combine hierarchically into
larger, less cohesive units \cite{ravasz2002,vazquez2004}. The
hierarchical scale-free model predicts that highly abundant network
motifs always aggregate into larger motif clusters centred around
network hubs in order to distribute a large number of motifs over a
comparatively small number of nodes \cite{vazquez2004}.

\begin{figure}
  \centering
  \includegraphics[width=\linewidth]{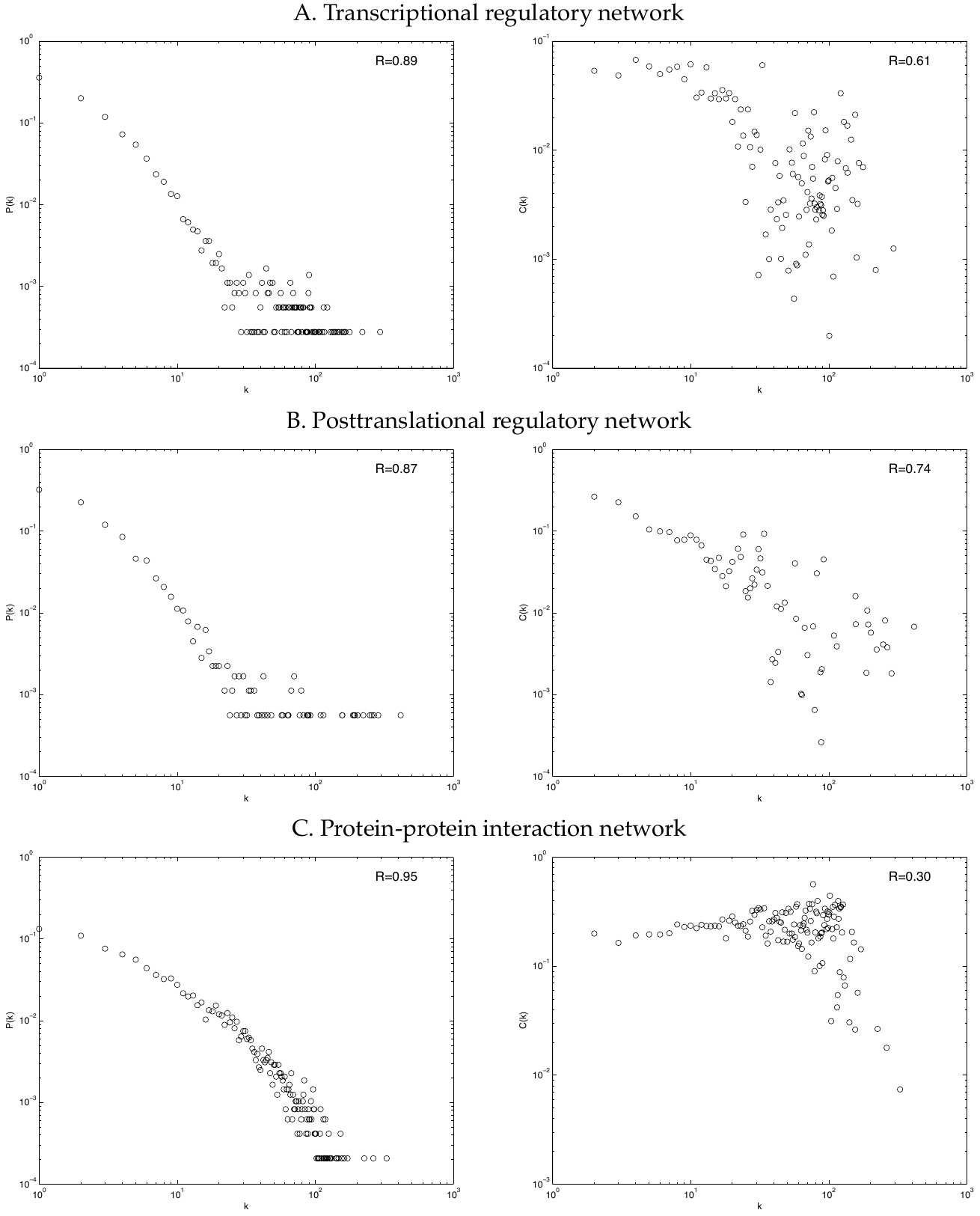}
  \caption{Degree distribution $P(k)$ (left) and clustering
    coefficient $C(k)$ (right) as a function of the degree $k$ for
    three molecular interaction networks in yeast. In the upper right
    corner of each figure, the correlation coefficient $R$ of the data
    to the best power-law fit is given.}
  \label{fig:degree-cc}
\end{figure}

Although \citet{vazquez2004} showed that several biological networks
could be approximated by the hierarchical scale-free model, new data
on these networks has accumulated since then. We calculated the degree
distribution $P(k)$ and clustering coefficient distribution $C(k)$ for
the transcriptional \cite{harbison2004} and posttranslational
\cite{ptacek2005} regulatory networks as well as for the physical
protein-protein interaction network \cite{stark2006} in yeast (Figure
\ref{fig:degree-cc}). For all three networks the degree distribution
fits well to a power-law (correlation coefficient $R = 0.89, 0.87,
0.95$ respectively), with deviations mainly at the (relatively few)
high-degree nodes or hubs. The clustering coefficient distribution
however shows significant deviation from power-law scaling already at
medium-degree nodes ($R=0.61, 0.74, 0.30$ respectively). In other
words, many medium-degree nodes in these networks have a significantly
higher clustering coefficient than expected from the hierarchical
scale-free model, suggesting the presence of an additional
organizational level.

\subsection{A network motif aggregation statistic to quantify local
  modularity}
\label{sec:netw-motif-clust}

We hypothesized that deviations of the clustering coefficient
distributions from power-law behavior are due to a local aggregation
of network motifs around specific nodes which are not necessarily
hubs.  To quantify this aggregation of network motifs, we made the
following considerations. If two networks share the same number of
instances of a given motif, then the motif is more aggregating in the
network where fewer nodes participate in one of the motif
instances. Conversely, if two networks have the same number of nodes
participating in motif instances, the motif is more aggregating in the
network which has the highest number of motif instances among these
nodes. We defined a network motif aggregation statistic $\S$ for any
three-node motif, having exactly these properties, as the ratio
\begin{equation}\label{eq:5}
  \S = \frac{N}{\sqrt{n_1 n_2 n_3}}
\end{equation}
where $N$ is the total number of motif instances and $n_1$, $n_2$ and
$n_3$ are the number of network nodes which participate at least once
in a motif at each of the three possible motif nodes (see Methods for
details).

We used this statistic to compare the real networks to randomized
networks which preserve the in- and out-degree distributions as well
as the total number of instances of the input motif of the real
network. This random network ensemble is different from the usual one
which only preserves the degree distributions. By adding the
constraint to also preserve the total motif count, we ensure to assess
aggregation of network motifs independent of their abundance or
enrichment.  We say a network exhibits \emph{local modularity} (as
opposed to hierarchical modularity) with respect to a certain motif if
its aggregation statistic is significantly higher in the real network
than in the randomized networks.

\subsection{Feedforward loop aggregation in yeast regulatory networks
  is independent of enrichment}
\label{sec:netw-motif-aggr}

We first considered the feedforward loop (FFL) in the transcriptional
and posttranslational regulatory network. The transcriptional FFL is
undoubtedly the best studied network motif and its functional role and
aggregation have been described in several studies
\cite{lee2002,shen-orr2002,kashtan2004,kalir2004,dobrin2004,zhang2005}.
It is strongly enriched ($P<0.001$) in our network and also
significantly aggregating ($P<0.001$). Interestingly, the FFL is not
at all enriched in the posttranslational network ($P>0.999$), where it
in fact occurs significantly less often than expected by
chance. However the posttranslational FFL is strongly aggregating
($P<0.001$). This result already indicates that network motif
enrichment and aggregation are not in one-to-one relation like in the
hierarchical scale free model and that the deviations of the
clustering coefficient distributions from the hierarchical scale-free
model in both networks (Figure \ref{fig:degree-cc}) are well
represented by the network motif aggregation statistic.

\subsection{Composite network motifs also exhibit local modularity
  independent of their enrichment}
\label{sec:comp-netw-motifs}

\begin{figure*}
  \centering
  \includegraphics[width=\linewidth]{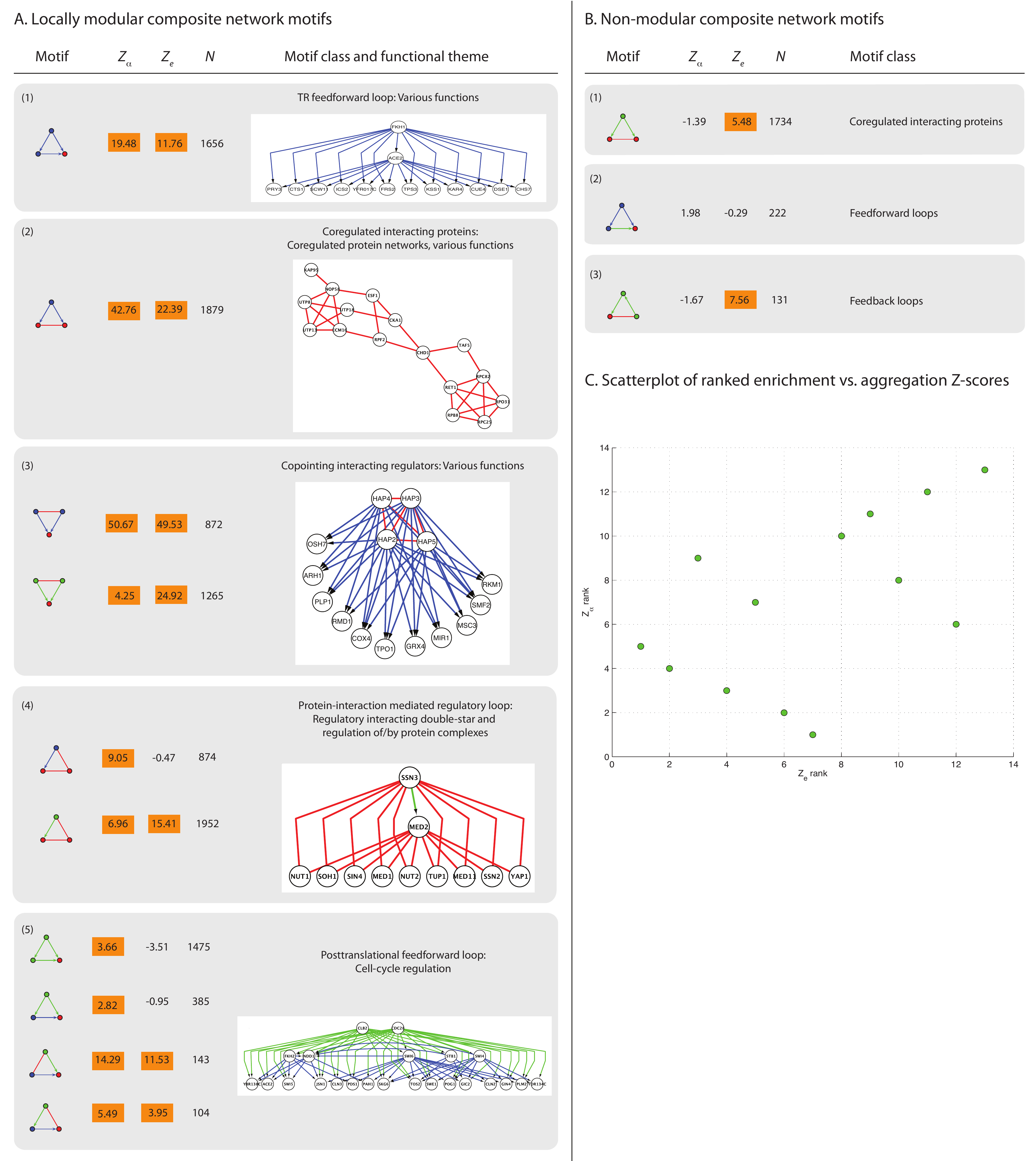}
  \caption{Significant locally modular (\textbf{A}) and non-modular
    (\textbf{B}) composite network motifs in the integrated yeast
    network, organized by common motif classes and functional
    themes. Shown is for each motif the aggregation $Z$-score
    ($Z_\alpha$), the enrichment $Z$-score ($Z_e$) and the number of
    instances ($N$), and for each functional theme an example of a
    motif cluster. Significant $Z$-scores ($P<0.005$) are highlighted
    in orange. Interaction color legend: TR, PTL and PPI interactions
    respectively in blue, green and red. \textbf{C.}  Scatterplot of
    $Z_e$ \textit{vs.} $Z_\alpha$ ranks (in ascending order) for all
    motifs in panel \textbf{A} and \textbf{B}.}
  \label{fig:aggr-motifs}
\end{figure*}

An additional reason why the clustering coefficient distributions may
deviate from the hierarchical scale-free model is the fact that the
transcriptional, posttranslational and protein-protein interaction
network do not exist in isolation but are intertwined with each other.
Network motifs composed of multiple interaction types represent the
functional relationships between different levels of regulation in a
cell \cite{yeger-lotem2004,yu2006,zhang2005}.  Hence we examined the
enrichment and aggregation of all three-node composite motifs which
occur at least 100 times in the integrated network of transcriptional,
posttranslational and protein-protein interactions (Figure
\ref{fig:aggr-motifs}). There appears to be no strong relation between
network motif enrichment and aggregation and in particular, there are
several examples of motifs which are not enriched yet display
significant aggregation (Figure \ref{fig:aggr-motifs}A). The Spearman
rank correlation between the enrichment and aggregation $Z$-scores is
$0.55$, indicating that both properties are at best weakly correlated
to each other (Figure \ref{fig:aggr-motifs}C).

\subsection{An algorithm to identify local clusters of overlapping
  network motifs}
\label{sec:netw-motif-clust-1}

\begin{figure}
  \centering
  \includegraphics[width=\linewidth]{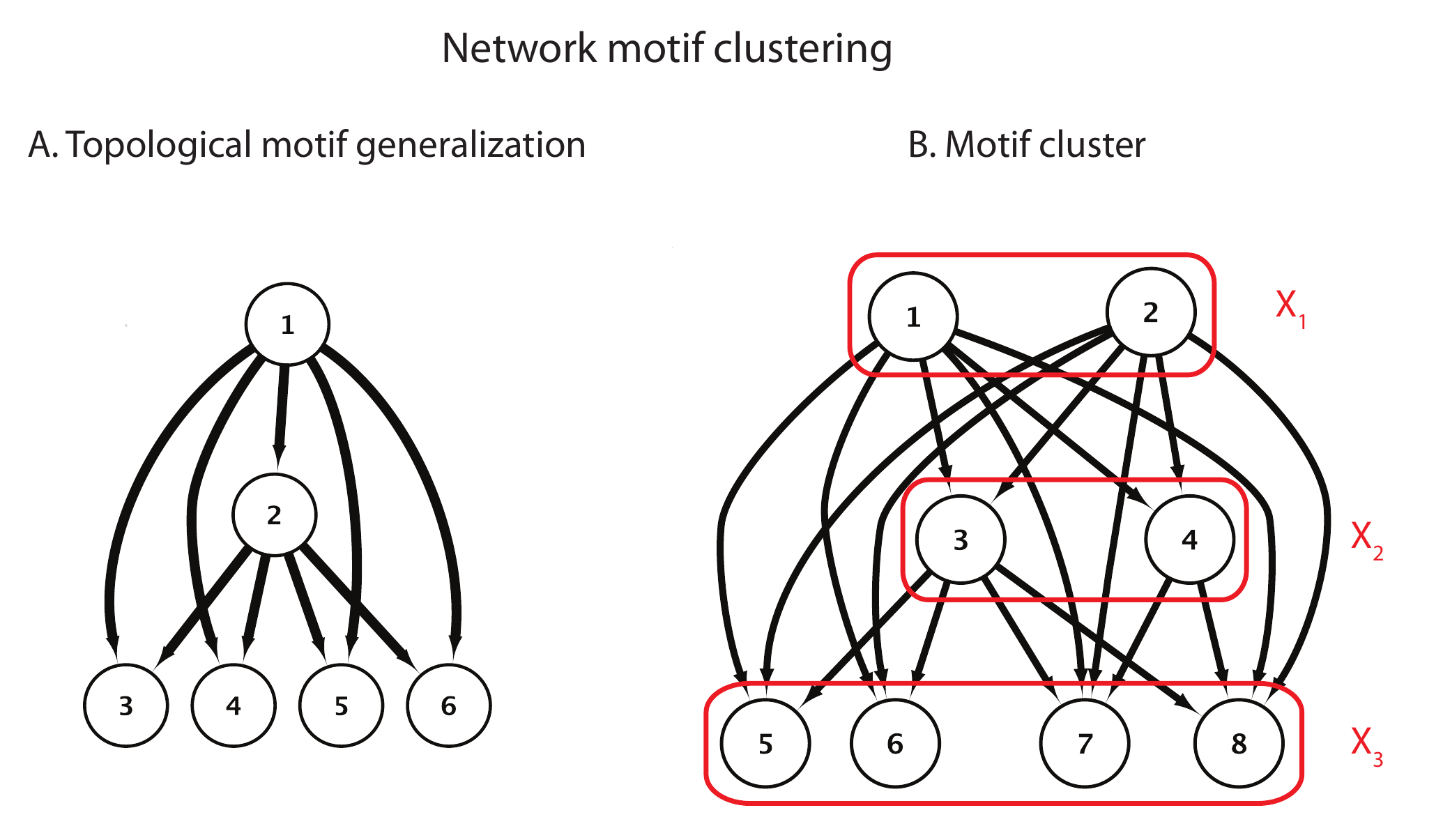}
  \caption{\textbf{A.} Example of a topological motif generalization
    where all possible motif instances (in this case FFL) between the
    nodes are present. \textbf{B.} Example of a motif cluster (in this
    case FFL) with a high aggregation score (high number of motif
    instances relative to the number of nodes in $X_1$, $X_2$ and
    $X_3$).}
  \label{fig:motif-cluster}
\end{figure}

To analyze in more detail the functional role of network motif
aggregation, we developed an algorithm to identify network motif
clusters.  \citet{kashtan2004} introduced the concept of topological
motif generalizations which consist of perfect motif replications
along one of the motif nodes (Figure \ref{fig:motif-cluster}A). To
allow for the possibility of imperfect networks with missing
interactions, we further generalized this concept and defined motif
clusters as subnetworks which locally maximize the aggregation
statistic $\S$ (cfr. eq. (\ref{eq:5})).  In a motif cluster, each
motif node $i$ corresponds to a `node role' \cite{kashtan2004} and is
replicated into a set of cluster nodes $X_i$ (Figure
\ref{fig:motif-cluster}B). The aggregation score of a cluster is
defined as the aggregation statistic restricted to the subnetwork
formed by $X_1$, $X_2$ and $X_3$.  To find high-scoring clusters, we
defined cluster membership weights for each node role, similar to
spectral weights for matrices \cite{newman2006}, such as the PageRank
\cite{brin1998} or hub- and authority \cite{kleinberg1999} weights: a
node gets a high weight in role 1, if it belongs to many motif
instances together with nodes which have high weights in role 2 and 3,
and similarly for the other roles. This yields a set of multilinear
equations in the membership weights for each role which are easily
solved numerically. After taking a suitable threshold on the weight
vectors a high-scoring cluster is obtained. The algorithm continues in
an iterative fashion by removing from the network all motif instances
assigned to the previous cluster and repeating the procedure until no
more instances remain.  Since motif instances are partitioned, nodes
and edges can belong to multiple clusters. We refer to the Methods
section for more details.

\subsection{Comparison with \citeauthor{zhang2005}}
\label{sec:comp-with-zhang}

\citet{zhang2005} defined network themes as classes of motif clusters
based on visual inspection of composite network motifs. Our definition
of local modularity on the other hand is based on the significance of
the network motif aggregation statistic and provides an unbiased and
rigorous method for identifying network themes. Three of the network
themes discovered by \citet{zhang2005} pertain to the networks studied
here, namely the transcriptional feedforward loop (Figure
\ref{fig:aggr-motifs}A(1)), the transcriptionally coregulated
interacting proteins motif (Figure \ref{fig:aggr-motifs}A(2)) and the
copointing interacting transcription factors motif (Figure
\ref{fig:aggr-motifs}A(3)). In all three cases our method found a
highly significant aggregation $Z$-score, confirming the validity of
our approach.

Having furthermore an automated clustering algorithm allows to
identify functional themes associated to each locally modular network
motif. For instance, among the functional categories enriched in
transcriptional FFL clusters, we find mainly the core processes
associated with transcription such as transcriptional control, DNA
binding and regulation of metabolic processes (Supplementary Table
S1), supporting the hypothesis that transcriptional FFLs play a
universal information-processing role\cite{alon2007b}. For the
transcriptionally coregulated interacting proteins motif, it is
usually assumed that enrichment and clustering reflects a `regulonic
complex' theme in which transcriptionally coregulated interacting
proteins are often members of a protein complex
\cite{zhang2005,yeger-lotem2004,tan2007}. We found that high-scoring
coregulated protein clusters sometimes overlap with known protein
complexes (Supplementary Table S2), but more often form `functional
protein networks' \cite{yosef2009} (Figure \ref{fig:aggr-motifs}A(2)):
subnetworks of the PPI network enriched for a particular function and
identified by overlaying the protein interaction network with an
additional layer of information, in this case regulator-target
data. Functional coregulated protein networks can be of practical
interest to generate detailed hypotheses for the different functions
in which a particular regulator is involved. For instance ABF1 is a
multifunctional global regulator, but its set of targets in the
transcriptional network is only enriched for tRNA synthesis. Network
motif clustering on the other hand identifies protein networks
regulated by ABF1 enriched for several more categories, many of which
are consistent with current knowledge, such as general DNA binding
function, regulation of ribosome biosynthesis \cite{planta1995},
nuclear transport \cite{loch2004}, etc. (Supplementary Table
S3). Interestingly, the posttranslationally coregulated interacting
proteins motif is not significantly aggregating, although it is
significantly enriched (Figure \ref{fig:aggr-motifs}B(1)). This means
that targets regulated by the same kinase physically interact more
often than randomly selected proteins, but the resulting coregulated
protein networks are not more dense than expected by chance. This
result is consistent with the fact that protein complexes can be
posttranslationally regulated by regulating just one instead of all of
its components (see also Section \ref{sec:regul-ofby-prot}).

Transcription factors often function as a complex and the binding
sites for these transcription factors occur more frequently within the
same promoter regions \cite{harbison2004} leading to a `copointing'
theme \cite{zhang2005}. The copointing interacting regulators motif is
significantly aggregating at transcriptional and posttranslational
level (Figure \ref{fig:aggr-motifs}A(3)).  At the transcriptional
level, copointing interacting regulator pairs include well-known
co-operating transcription factors like the cell cycle transcription
factors SWI4--SWI6--MBP1, the ribosomal protein regulators RAP1--FHL1,
the galactose response regulatory complex GAL3--GAL80, and others. At
the posttranslational level, many clusters come from the
multi-functional cyclin-dependent kinases (CDK) CDC28 or PHO85
complexed with one of their many cyclin activators
\cite{mendenhall1998,huang2007}.

We conclude that highly abundant network motifs which are strongly
enriched as well as aggregating likely have a universal
information-processing role which extends across various functional
categories.

\subsection{Posttranslational feedforward loop aggregation reflects
  a cell-cycle regulation theme}
\label{sec:cell-cycle-regul}

Perhaps the most surprising finding in our analysis is the existence
of network motifs which aggregate significantly but are not enriched
(Figure \ref{fig:aggr-motifs}).  We hypothesize that local modularity
of a network motif without significant enrichment indicates that this
motif is important for specific biological functions but does not play
a universal role like the strongly enriched motifs in the previous
section. Two examples of such motifs are the posttranslationally
controlled feedforward loops (Figure \ref{fig:aggr-motifs}A(5)).

More than half of the posttranslational FFLs and mixed
posttranslational-transcriptional FFLs belong to clusters regulated by
the multi-functional CDKs CDC28 or PHO85 and these clusters are indeed
often enriched for cell-cycle related functions (Supplementary Tables
S4 and S5). CDC28 is the central coordinator of the yeast cell cycle
\cite{mendenhall1998} and it has been shown that the mixed
posttranslational-transcriptional FFLs regulated by it are important
transducers between cell cycle regulatory signals and responses, using
dynamical models for individual motif instances
\cite{csikasz-nagy2009}. Our approach on the other hand reveals the
overlapping nature of these motifs.  For instance, cluster 1 (depicted
in Figure \ref{fig:aggr-motifs}A(5)) contains three transcription
factors functioning in G1/S transition (SWI4, SWI6, STB1), one in G2
phase (FKH2) and one in G2/M transition (NDD1).  The complexity of the
overlapping motif structure is further emphasized by the fact that
NDD1 not only functions as a transcriptional transducer of the cell
cycle signal, but also as a response target of the four other
transcription factors. The target proteins in this cluster are
enriched for several cell-cycle related functions (Supplementary Table
S5) and eleven of the sixteen targets are periodically expressed
\cite{spellman1998,delichtenberg2005}.  PHO85 is another CDK with a
multifunctional role in cell cycle control and other processes
\cite{huang2007}. Like CDC28 it is activated by a large family of
cyclins. The transcription factors associated to clusters regulated by
PHO85 contain the PHO85 substrates PHO4, GCN4 and SWI5 whose
phosphorylation is important for the role of PHO85 in regulating
environmental signalling response and the cell cycle \cite{huang2007}.
This suggests that the posttranslational-transcriptional FFL plays a
similar dynamical role in transducing PHO85 regulatory signals as for
CDC28 \cite{csikasz-nagy2009}.

The cell-cycle is a complex process and cell-cycle kinases often also
interact physically with their target substrates. As a result there
are two motifs involving all three interaction types (Figure
\ref{fig:aggr-motifs}A(5)) which almost all overlap with mixed
posttranslational-transcriptional FFLs. The aggregation significance
is consistent across all four posttranslational FFLs but the
enrichment is not (Figure \ref{fig:aggr-motifs}A(5)). This is not in
contradiction with the previous result that enriched locally modular
motifs play a universal role, since the vast majority of
triple-interaction motifs involve cell-cycle regulators,
i.e. `universal' always refers to the network at hand.

In summary, we can say that the posttranslational regulatory network
exhibits an overall lack of feedforward loops, presumably because it
operates on a much shorter timescale than the transcriptional
regulatory network to elicit fast information-processing responses,
typically in the form of signaling cascades
\cite{alon2007b}. Posttranslational feedforward loops (pure as well as
composite) do seem to play an important role however in regulation of
the cell-cycle, and this `local' role is reflected in a significant
aggregation of these motifs around the core CDKs CDC28 and PHO85.

\subsection{Transcriptional and posttranslational protein
  interaction-mediated regulatory loop aggregation reflects a
  regulatory protein complex theme}
\label{sec:regul-ofby-prot}

Another motif which is not enriched yet displays significant
aggregation is the protein-interaction mediated transcriptional
regulatory loop (Figure \ref{fig:aggr-motifs}A(4)), a circuit that is
thought to serve for feedback mechanisms between a regulator-target
pair via a common partner in the protein interaction network
\cite{yeger-lotem2003}.  In the equivalent posttranslational motif all
interactions can occur simultaneously and its proposed function is
that of a `scaffold motif' where the biochemical interaction between
the regulator and its target substrate is enabled by the common
interactor \cite{ptacek2005}.  A natural cluster generalization of
such feedback or scaffold circuits is a `regulonic star', where
multiple targets of a regulator (`spokes') interact with the same
feedback or scaffold mediator (`hub').  Our algorithm identified
several such modules as high-scoring motif clusters (Supplementary
Table S6 and S7).  For instance transcriptional cluster 11 consists of
ABF1, a DNA binding protein that regulates multiple nuclear events,
regulating a set of ten nuclear transport genes which all interact
with PSE1, a nuclear transport receptor which also interacts with
ABF1. A link between ABF1 and the nuclear transport machinery via PSE1
is known \cite{loch2004}.  Transcriptional cluster 14 has HSP82 as the
hub protein.  HSP82 is one of two yeast genes encoding for HSP90, a
protein folding chaperone which plays a central role in various
aspects of cellular signaling \cite{mcclellan2007,zhao2005}.  Binding
of HSP90 to HAP1, the regulator of cluster 14, is necessary for heme
activation of HAP1 \cite{lee2003}.  This cluster may represent a
feedback mechanism since TAH1, a cofactor of HSP90 \cite{zhao2005}, is
one of its spoke proteins. Posttranslational cluster 43 is an example
of a scaffolding regulonic star. It consists of the mitotic B-type
cyclin CLB2 which phosphorylates nine proteins involved in budding,
cell polarity and filament formation, which all interact with NAP1, a
protein which is known to interact with and facilitate the function of
CLB2 \cite{kellog1995}.

We also found a relation between the protein-interaction mediated
regulatory loops and protein complexes in the form of a `regulatory
interacting (RI) double-star' cluster type, consisting of one or a few
regulator--target pairs which share a common set of partners in the
protein interaction network. Usually the spoke proteins in such a RI
double-star mutually interact and form the components of a protein
complex, often together with the hub protein (Supplementary Tables S8
and S9).  For instance, in posttranslational cluster 32, SSN3 (also
called SRB10) phosphorylates MED2, a component of the RNA polymerase
II Mediator complex, and both interact with eight other Mediator
components and the transcription factor YAP1 (depicted in Figure
\ref{fig:aggr-motifs}A(4)).  It is known that posttranslational
modification of Mediator components affects its function, and that
SRB10 is part of a module whose binding to the Mediator complex
determines if Mediator can associate with pol II or not
\cite{bjoerklund2005}.  YAP1, a bZIP transcription factor required for
oxidative stress tolerance, is related to the Mediator complex via a
transcriptional RI double-star in which 25 components of the Mediator
complex interact with YAP1 and SRB6 and SRB7, two other components of
the Mediator complex that are transcriptionally regulated by YAP1.
The Mediator complex acts as a bridge between gene-specific
transcription factors and the basal pol II transcription machinery
\cite{bjoerklund2005} and mutants of the general transcription factor
TFIIA are unable to grow in conditions that require the oxidative
stress response \cite{kraemer2006}.  Another example of a
transcriptional RI double-star is transcriptional cluster 20 where
HAP4 regulates GCN4 and both interact with five components of the
SWI/SNF complex, which regulates transcription by nucleosome
remodeling. HAP4 and GCN4 are two transcriptional activators which
target SWI/SNF to appropriate promoters \cite{neely1999}.

The protein complexes which appear in transcriptional regulatory
interacting double-stars are all regulatory complexes involved in the
different steps from chromatin remodelling to transcription,
translation and posttranslational control (Supplementary Table S8).
In cases where the hub protein also belongs to the complex, this
suggests that the RI double-star acts like a two-node feedback loop in
which the transcription factor regulates the complex by regulating one
or a few of its components, and in turn the complex regulates the
transcription factor by interacting with it at the protein level.  A
composite feedback mechanism using a slow (transcriptional) and fast
(protein-protein interaction) timescale is known to enhance stability
around a steady state \cite{alon2007b}. For the posttranslational
motif on the other hand, RI double-star clusters reflect regulation of
the protein complex, and hence we find a much broader range of protein
complexes besides regulatory complexes (Supplementary Table S9). This
more universal role is again consistent with the fact that the
posttranslational motif is strongly enriched but the transcriptional
is not. The interpretation of a posttranslational RI double-star
cluster is that the kinase and protein complex form a two-component
loop where the complex acts as a scaffold protein for its own
regulation.

\begin{figure*}
  \centering
  \includegraphics[width=\linewidth]{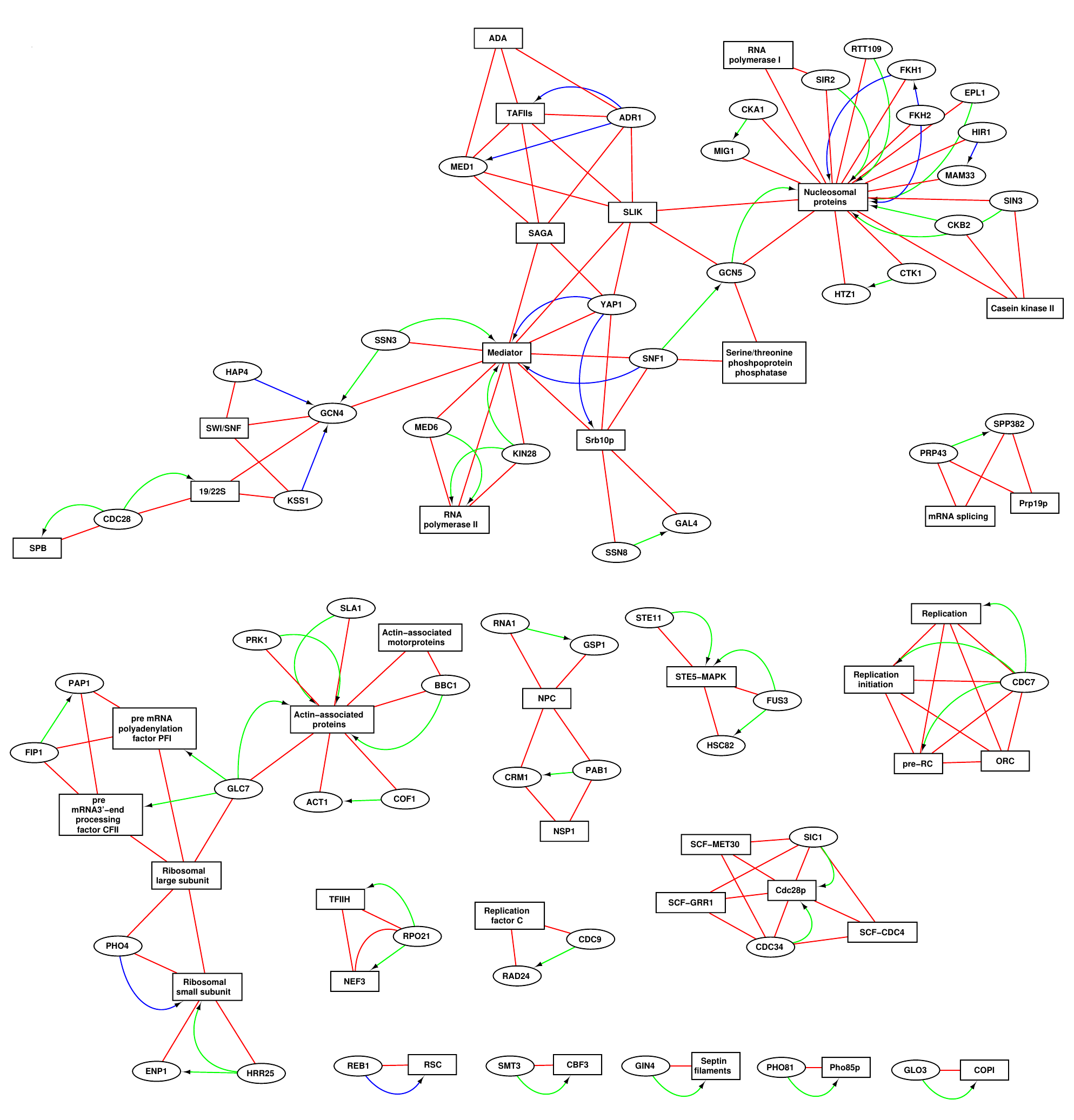}
  \caption{Global map of protein complex regulation through
    transcriptional and posttranslational regulatory interacting
    double-star clusters. Oval nodes are proteins, rectangular nodes
    are protein complexes which overlap significantly ($P<10^{-4}$)
    with a RI double-star cluster. Two-node mixed feedback loops
    indicate that the target hub protein belongs to the same complex
    as the spoke nodes. This figure is based on the data in
    Supplementary Tables S8 and S9. The interaction colors are the
    same as in Figure \ref{fig:aggr-motifs}.}
  \label{fig:ristar-map}
\end{figure*}

Regulatory interacting double-star clustering of protein-interaction
mediated regulatory loops induces a higher-level, global map of
protein complex regulation. In Figure \ref{fig:ristar-map} we
considered all protein complexes which overlap significantly with RI
double-star clusters with at least three (transcriptional) or four
(posttranslational) spoke proteins. This map shows a high amount of
two-component regulator--complex feedback loops, with a central role
for the Mediator complex and the nucleosomal proteins.  Interestingly,
transcriptional and posttranslational regulation are heavily
intertwined in this map. Some complexes (Mediator, small ribosomal
subunit, nucleosomal proteins) are regulated by transcriptional as
well as posttranslational RI double stars, while others (Srb10p, SLIK)
play a feedback or scaffolding role for transcriptional as well as
posttranslational regulatory interactions.  Figure
\ref{fig:ristar-map} provides a novel kind of coarse grained
integrated network representation which complements previous thematic
maps of compensatory and regulonic complexes \cite{zhang2005}.

The protein-interaction mediated transcriptional regulatory loop has
previously been found enriched \cite{zhang2005} or not enriched
\cite{yeger-lotem2004} in different datasets. We calculated the
aggregation statistic in these two datasets as well as for a network
of literature-curated protein-protein interactions \cite{reguly2006},
in which the motif is also enriched, and found in all cases a
consistent statistically significant aggregation (data not shown). The
protein interaction networks where the motif is enriched are targeted
towards co-complex interactions, while the networks were it is not
enriched also contain interactions derived from yeast two-hybrid
studies.  Thus we find that network motif aggregation is a more robust
property than network motif enrichment and that the enrichment in
co-complex based networks can be explained by the relation between the
protein-interaction mediated transcriptional regulatory loop and
protein complexes as detailed in the previous paragraphs. Our results
also highlight the advantage of using an unbiased clustering approach
to identify network themes, since \citet{zhang2005} did consider the
transcriptional protein interaction-mediated loop but found no theme
or cluster generalization for it.

\section{Conclusions}

Network motifs form the basic building blocks of complex networks and
previous studies have suggested an intimate relation between the
enrichment of a motif and its tendency to aggregate into functional
modules. Evolutionary studies hypothesized that network motif clusters
have evolved by simple duplication-divergence mechanisms, concluding
that the abundance of a motif is merely a by-product of the emergence
of these aggregated motif structures. On the other hand, a purely
topological analysis in hierarchical scale-free random networks has
shown that the only way to distribute a large number of motifs over a
comparatively small number of nodes is to aggregate the motifs around
the network hubs, indicating that motif aggregation follows from motif
enrichment.

Here we introduced a novel method for assessing in a quantitave way
the statistical significance of network motif aggregation. Using an
integrated network of transcriptional, posttranslational and
protein-protein interactions in yeast, we showed that our method
produces results which are consistent with deviations of the
clustering coefficient distribution from the hierarchical scale-free
model and correctly recapitulate previous findings. We showed
furthermore that significant aggregation reflects local modularity of
network motifs around nodes which are not necessarily hubs in the
total network but play a core role in specific biological processes.
Using a novel network motif clustering algorithm we found that if in
addition to aggregate significantly, a motif is also enriched, it
likely plays a universal functional role across many biological
processes. If the motif aggregates significantly without enrichment,
it is likely specific for one or a few processes.  We identified novel
functional network themes for such motifs, like a cell-cycle
regulatory theme for posttranslationally controlled feedforward loops
and a regulation of or by protein complexes for protein-interaction
mediated regulatory loops.

Our results show that network motif aggregation is an important
organizational principle of molecular interaction networks which is
independent of network motif enrichment and in particular is more
robust against using heterogeneous experimental sources of interaction
data and less sensitive to the incompleteness of all current
interaction networks. We hypothesize that network motif aggregation is
the more fundamental organizational principle and that network motif
enrichment may follow from aggregation if the local modularity
property extends across a sufficiently large fraction of biological
processes represented in the network at hand.  Future work to support
this hypothesis should be directed towards comparative analyses to
confirm these results across multiple organisms and towards improving
the current duplication-divergence models for network growth and
hierarchical scale-free models for network topology in order to
reproduce and understand the evolutionary origin of the complex
modular organization observed in the integrated yeast interaction
network.

\section{Methods}
\label{sec:methods}

\subsection{Network data}

The protein-protein interaction network ($36391$ interactions between
$4847$ proteins) was extracted from the BioGRID database
\cite{stark2006} (\url{http://www.thebiogrid.org}). The
transcriptional regulatory network ($11373$ interactions between $198$
transcription factors and $3535$ target genes) was obtained from
\citet{harbison2004} (\url{http://fraenkel.mit.edu/Harbison/}), using
a $P$-value cutoff of $0.005$. The posttranslational regulatory
network ($5630$ interactions between $264$ regulators and $1653$
target proteins) was extracted from the BioGRID database (interactions
annotated as `biochemical activity') \cite{stark2006}.  The bulk of
this network (4621 interactions) comes from the phosphorylation
network of \citet{ptacek2005}
(\url{http://networks.gersteinlab.org/phosphorylome/}).
Self-interactions were not kept in any of the networks.

\subsection{Network motif aggregation statistic}
\label{sec:netw-motif-aggr-2}

For a given 3-node motif we choose a particular labeling of its nodes
and define its aggregation statistic as
\begin{equation}\label{eq:6}
  \S = \frac{N}{\sqrt{n_1 n_2 n_3}}
\end{equation}
where $N$ is the total number of motif instances and $n_1$, $n_2$ and
$n_3$ are the number of nodes which participate at least once in a
motif in node role $1$, $2$ and $3$, respectively. $\S$ has the
intuitive properties that it is higher (more aggregation) in a network
with a fixed number of motif instances distributed over a smaller
number of nodes or in a network with more motif instances distributed
over a fixed number of nodes. We have $N\leq n_1 n_2 n_3$, and the
maximum is attained for perfect topological motif generalizations as
in Figure \ref{fig:motif-cluster}A where all possible motif instances
are indeed present. The square root in eq. (\ref{eq:6}) ensures that
$\S$ will be higher for bigger topological motif generalizations as
well as for larger sets of nodes with a significant number of motif
instances between them (as in Figure \ref{fig:motif-cluster}B), and is
thus suitable to measure motif aggregation in noisy interaction data
with potentially a large number of missing interactions.

\subsection{Local network motif aggregation score}
\label{sec:local-network-motif}

We define network motif clusters as subnetworks which locally maximize
the network motif aggregation statistic. To make this precise, for a
given 3-node input motif, we define a 3-dimensional motif array $T$ by
\begin{align*}
  T_{ijk}=
  \begin{cases}
    1 & \text{ if there exists a motif instance between $i$, $j$, $k$}\\
    0 & \text{ otherwise}
  \end{cases}
\end{align*}
where $(i,j,k)$ is any triple of nodes and the order of the indices
corresponds to a particular labeling of the nodes in the motif.  $T$
can be constructed from the adjacency matrices of the networks
defining the motif. For instance, the motif array for the feedforward
loop in a directed network with adjacency matrix $A$ is given by
\begin{align*}
  T_{ijk} = A_{ij} A_{jk} A_{ik}.
\end{align*}
A motif cluster is now defined by three sets of nodes $(X_1,X_2,X_3)$
(cfr. Figure \ref{fig:motif-cluster}) and its aggregation score can be
written as
\begin{equation}\label{eq:2}
  \S(X_1,X_2,X_3) = \frac{\sum_{i\in X_1, j\in X_2, k\in X_3}
    T_{ijk}}{\sqrt{|X_1|\,|X_2|\,|X_3|}},
\end{equation}
where $|X|$ denotes the number of nodes in $X$.

\subsection{Network motif clustering algorithm}

We want to find $(X_1,X_2,X_3)$ which maximize the local aggregation
score.  To this end, we first find the best rank-1 approximation to
$T$ \cite{delathauwer2000}, i.e. find real-valued vectors $(u,v,w)$
maximizing
\begin{equation}\label{eq:3}
  \R(u,v,w) = \frac{\sum_{ijk}T_{ijk} u_iv_jw_k}{\|u\| \|v\| \|w\|},
\end{equation}
where $\|u\|=\sqrt{\sum_i u_i^2}$ is the length of $u$.  These
maximizing vectors can be found efficiently by a multilinear power
method \cite{delathauwer2000}. For a set of nodes $X$ we define an
index vector $u_X$ by
\begin{align*}
  u_{X,i}=
  \begin{cases}
    1 & \text{ if $i\in X$}\\
    0 & \text{ otherwise}
  \end{cases}
\end{align*}
such that $\S(X_1,X_2,X_3)=\R(u_{X_1},u_{X_2},u_{X_3})$. This property
is used to prove that for any $X_1$, $X_2$, $X_3$
\begin{multline}\label{eq:1}
  \bigl| \S_{\max} - \S(X_1,X_2,X_3)\bigr| \\
  \leq \sqrt{2}\R_{\max}\bigl( \|u-u_{X_1}\| + \|v-u_{X_2}\| +
  \|w-u_{X_3}\|\bigr)
\end{multline}
where $\S_{\max}$ is the (unknown) maximal value of $\S$ over all
possible node sets, $\R_{\max}$ is the (known) maximal value of $\R$
over all real-valued vectors, $(u,v,w)$ is the (known) best rank-1
approximation to $T$, and all vectors on the r.h.s. are normalized to
length 1. Using the fact that $(u,v,w)$ have nonnegative entries, it
is trivial to find $(X_1,X_2,X_3)$ which minimize respectively
$\|u-u_{X_1}\|$, $\|v-u_{X_2}\|$ and $\|w-u_{X_3}\|$, or equivalently,
maximize $\langle u,u_{X_1}\rangle$, $\langle v,u_{X_2}\rangle$ and
$\langle w, u_{X_3}\rangle$, where
\begin{equation}\label{eq:4}
  \langle u,u_{X}\rangle = \frac1{\sqrt{|X|}}\sum_{i\in X} u_i
\end{equation}
is the overlap between $u$ and $u_{X}$.  By eq. (\ref{eq:1}), these
$(X_1,X_2,X_3)$ are the best possible approximation to the highest
scoring motif cluster, given our knowledge of the best rank-1
approximation to $T$. The r.h.s of eq.  (\ref{eq:1}) gives a precise
estimate on the quality of this approximation.  Next we remove from
the motif array $T$ all entries corresponding to the motif instances
in the highest scoring motif cluster.  The procedure is repeated for
this truncated motif array and iterated until no more non-zero
entries remain, thus obtaining a partition of all motif instances into
high-scoring motif clusters.

For symmetric motifs (such as e.g. the coregulated interacting
proteins motif, Figure \ref{fig:aggr-motifs}A(2)), the motif array is
symmetric for interchanging two indices, $T_{ijk}=T_{ikj}$ for all
triples $(i,j,k)$. The algorithm proceeds in the same way as before
but ensures that a symmetric maximizer of eq. (\ref{eq:3}) is found
with $v=w$, resulting in symmetric clusters with $X_2=X_3$.

\subsection{Network randomization algorithms}
\label{sec:netw-rand-algor}

To assess network motif enrichment we generated random networks with
the same incoming and outgoing degree distributions as the real
networks as follows. A directed network with $N_e$ edges can be
represented by two index vectors $\{I,J\}$ of length $N_e$ such that
the $k$th edge points from node $I_k$ to node $J_k$. We first
generated a random permutation $\pi(J)$ of the indices in $J$. The
network represented by $\{I,\pi(J)\}$ automatically has the same in-
and out-degrees as the original network, except there may be some
unwanted self-interactions, i.e. indices $k$ where $I_k=\pi(J)_k$. We
swap these $\pi(J)_k$ with a randomly chosen entry $\pi(J)_l$ until we
obtain a corrected permutation $\pi_c(J)$ without any
self-interactions and a corresponding randomized network
$\{I,\pi_c(J)\}$. Undirected networks are treated in the same way
except we choose $I_k<J_k$ for every $k$. After randomly permuting $J$
and correcting for self-interactions as before, we swap columns for
every $I_k>\pi(J)_k$ and correct for any duplicate edges by again
randomly swapping entries in $\pi(J)$. This randomization strategy is
easier to implement and runs faster than the conventional single edge
swapping strategies.

\begin{figure}
  \centering
  \includegraphics[width=\linewidth]{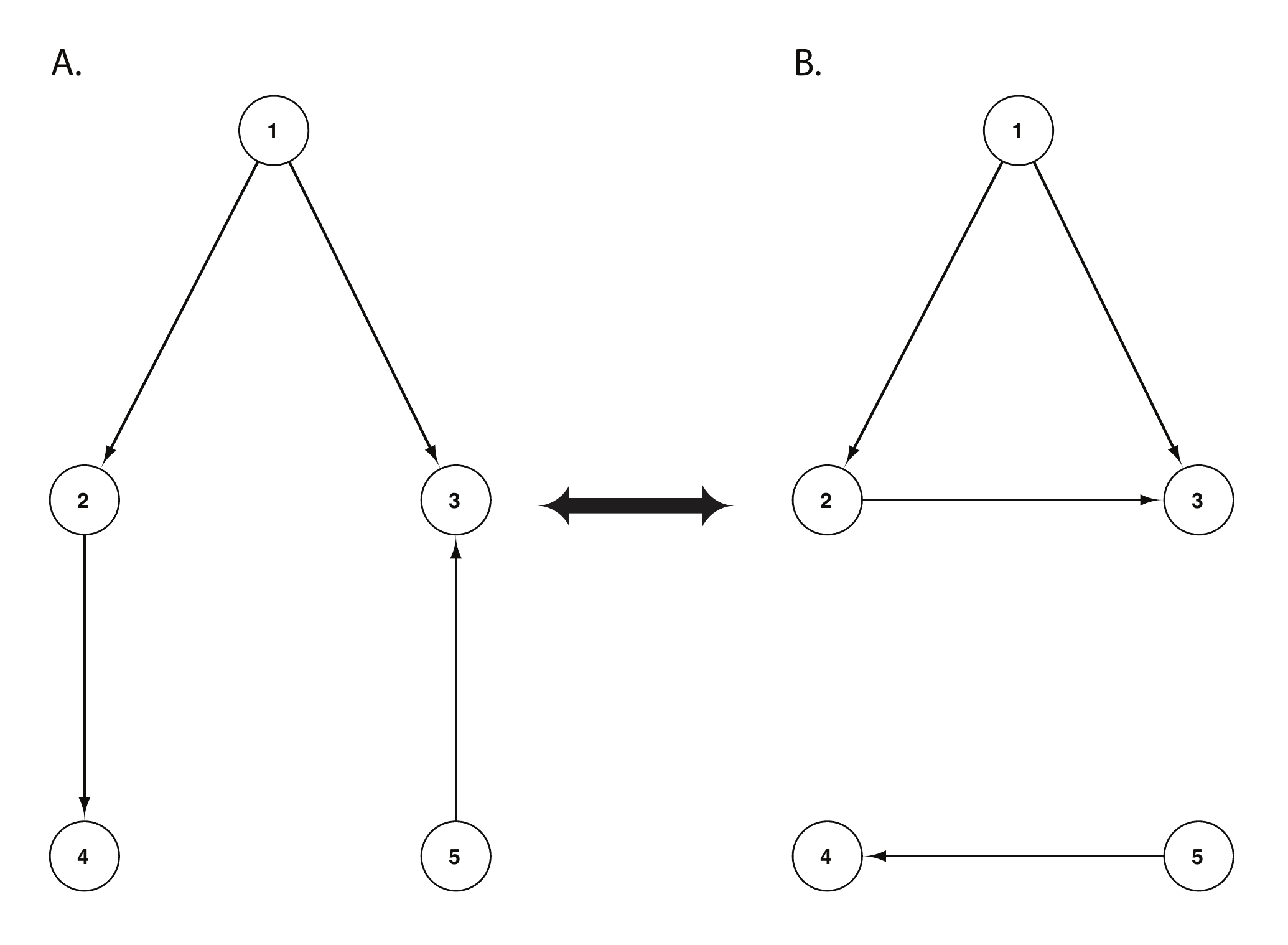}
  \caption{Example of edge swapping operations to increase (A$\to$B)
    or decrease (B$\to$A) the number of FFLs in a random network while
    keeping the in- and out-degree distributions constant..}
  \label{fig:swapedge}
\end{figure}

To assess network motif aggregation we generated random networks with
the same incoming and outgoing degree distributions as well as the
same total motif count as the real networks as follows. First we
generate random networks with the same in- and out-degrees as
described in the previous paragraph. If the total number of motif
instances is smaller in the random network than the real one, we
generate the list of all `incomplete' motif instances, triplets of
nodes with two motif interactions present and one absent (Figure
\ref{fig:swapedge}A). We randomly select one of these incomplete
motifs. For the two nodes with missing edge between them, we randomly
select an incoming, resp. outgoing edge. We then swap the endpoints of
these edges to `close' the incomplete motif while preserving the
degree distributions (Figure \ref{fig:swapedge}, A$\to$B). If the
total number of motif instances is larger in the random network than
the real one, we randomly select a motif instance and edge in that
motif (Figure \ref{fig:swapedge}B). We randomly select another edge
not belonging to any motif instance and swap the endpoints of the
motif edge with this edge to `open' a motif instance (Figure
\ref{fig:swapedge}, B$\to$A). We iterate between closing or opening
motif instances until the total number of motif instances is equal
between the random and real network.

\subsection{Network motif enrichment and aggregation significance}
\label{sec:netw-motif-enrichm}

To compute network motif enrichment significance, we generated 1000
random networks with the same in- and out-degree distributions as the
real transcriptional, posttranslational and protein-protein
interaction networks.  The enrichment $P$-value is defined as the
fraction of random networks having at least the same number of motif
instances as the real networks, and the $Z$-score is defined as
\begin{equation*}
  Z = \frac{N - \mu}{\sigma}
\end{equation*}
where $N$ is the number of motif instances in the real network and
$\mu$, resp. $\sigma$, is the mean, resp. standard deviation, of the
number of motif instances in the random network ensemble.  To compute
network motif aggregation significance, we generated for each input
motif 1000 random networks with the same in- and out-degree
distributions as well as the same total motif count as the
transcriptional, posttranslational and protein-protein interaction
networks. The aggregation $P$-value is defined as the fraction of
random networks having at least the same aggregation statistic as the
real network, and the $Z$-score is defined as in the previous
paragraph.  Notice that enrichment can be computed for all composite
motifs using a single ensemble of integrated random networks. On the
other hand, to compute aggregation, a separate random network ensemble
has to be generated for each input motif.

\subsection{Software}

A Network Motif Clustering Toolbox containing an implementation of the
network motif clustering algorithm as well as functions to generate
random networks and compute network motif enrichment and aggregation
significance is freely available for academic purposes, including
source code, from \url{http://omics.frias.uni-freiburg.de/software/}.
The toolbox operates under both Matlab
(\url{http://www.mathworks.com}) and Octave (\url{http://www.gnu.org/software/octave}).

\section{Acknowledgements}
\label{sec:acknowledgements}

We thank Eric Bonnet and Vanessa Vermeirssen for discussions and
testing the software. TM wishes to thank the Department of Mathematics
of the University of California, Davis for warm hospitality during
visits when part of this work was performed.  The work of TM, AJ and
YVdP was supported in part by IWT (SBO-BioFrame) and IUAP P6/25
(BioMaGNet).  The work of BN was supported in part by the National
Science Foundation under grant DMS-1009502.




\footnotesize{

\providecommand*{\mcitethebibliography}{\thebibliography}
\csname @ifundefined\endcsname{endmcitethebibliography}
{\let\endmcitethebibliography\endthebibliography}{}

}

\end{document}